\documentclass[conference]{IEEEtran}
\IEEEoverridecommandlockouts
\usepackage{cite}
\usepackage{amsmath,amssymb,amsfonts}
\usepackage{algorithmic}
\usepackage{graphicx}
\usepackage{textcomp}
\usepackage{xcolor}
\usepackage{gensymb}
\usepackage{subfig}
\usepackage[normalem]{ulem}
\useunder{\uline}{\ul}{}
\def\BibTeX{{\rm B\kern-.05em{\sc i\kern-.025em b}\kern-.08em
    T\kern-.1667em\lower.7ex\hbox{E}\kern-.125emX}}
    
\usepackage{etoolbox}
\makeatletter
\patchcmd{\@makecaption}
  {\scshape}
  {}
  {}
  {}
\makeatother
\begin{document}

\title{Adversarial Loss for Semantic Segmentation of Aerial Imagery\\
{}
}

\author{\IEEEauthorblockN{
Clint Sebastian, Raffaele Imbriaco, Egor Bondarev and Peter H.N. de With}

\IEEEauthorblockA{Department of Electrical Engineering,
Eindhoven University of Technology, 
Eindhoven, The Netherlands}
\IEEEauthorblockA{Email: \{c.sebastian, r.imbriaco, e.bondarev, p.h.n.de.with\}@tue.nl} 
}

\maketitle

\begin{abstract}
Automatic building extraction from aerial imagery has several applications in urban planning, disaster management, and change detection. In recent years, several works have adopted deep convolutional neural networks (CNNs) for building extraction, since they produce rich features that are invariant against lighting conditions, shadows, etc. Although several advances have been made, building extraction from aerial imagery still presents multiple challenges. Most of the deep learning segmentation methods optimize the per-pixel loss with respect to the ground truth without knowledge of the context. This often leads to imperfect outputs that may lead to missing or unrefined regions. In this work, we propose a novel loss function combining both adversarial and cross-entropy losses that learns to understand both local and global contexts for semantic segmentation. The newly proposed loss function deployed on the DeepLab~v3+ network obtains state-of-the-art results on the Massachusetts buildings dataset. The loss function improves the structure and refines the edges of buildings without requiring any of the commonly used post-processing methods, such as Conditional Random Fields. We also perform ablation studies to understand the impact of the adversarial loss. Finally, the proposed method achieves a relaxed $F_1$ score of 95.59\% on the Massachusetts buildings dataset compared to the previous best $F_1$ of 94.88\%.
\end{abstract}


\section{Introduction}
Several developments in the collection of remote sensing imagery have resulted into the availability of high-resolution aerial image datasets for exploring applications such as object detection, image retrieval, etc. Detection and recognition of objects in aerial imagery is crucial for urban planning, disaster mitigation, map making, and change detection. One of the most prominent objects that are maintained and updated are buildings. Therefore, building extraction reaps a plethora of benefits for the aforementioned applications. Because of the increasing amount of aerial imagery, automating the detection process becomes desirable. In recent years, advances of machine learning along with the development of low-cost hardware have resulted in high-performance object detection algorithms. However, building detection from remote sensing images still faces several challenges, where large variations in building appearance (varying building shapes, sizes, and colors), lighting conditions and shadows, often pose difficulties for reliable detection.

\begin{figure}[t]
\captionsetup[subfloat]{labelformat=empty}
    \centering
    \subfloat{\includegraphics[width=0.49\linewidth]{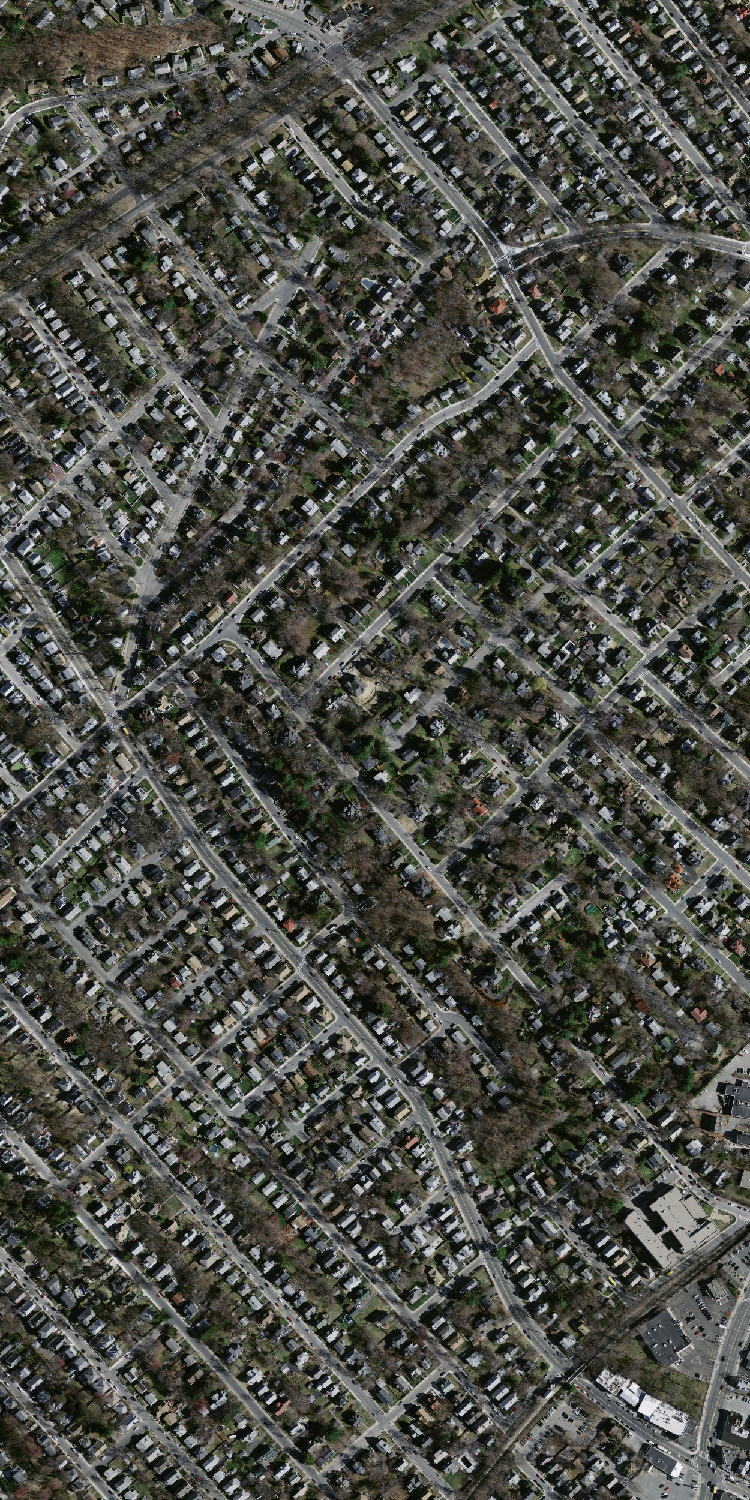}}
    \hspace{0.02mm}
    \subfloat{\includegraphics[width=0.49\linewidth]{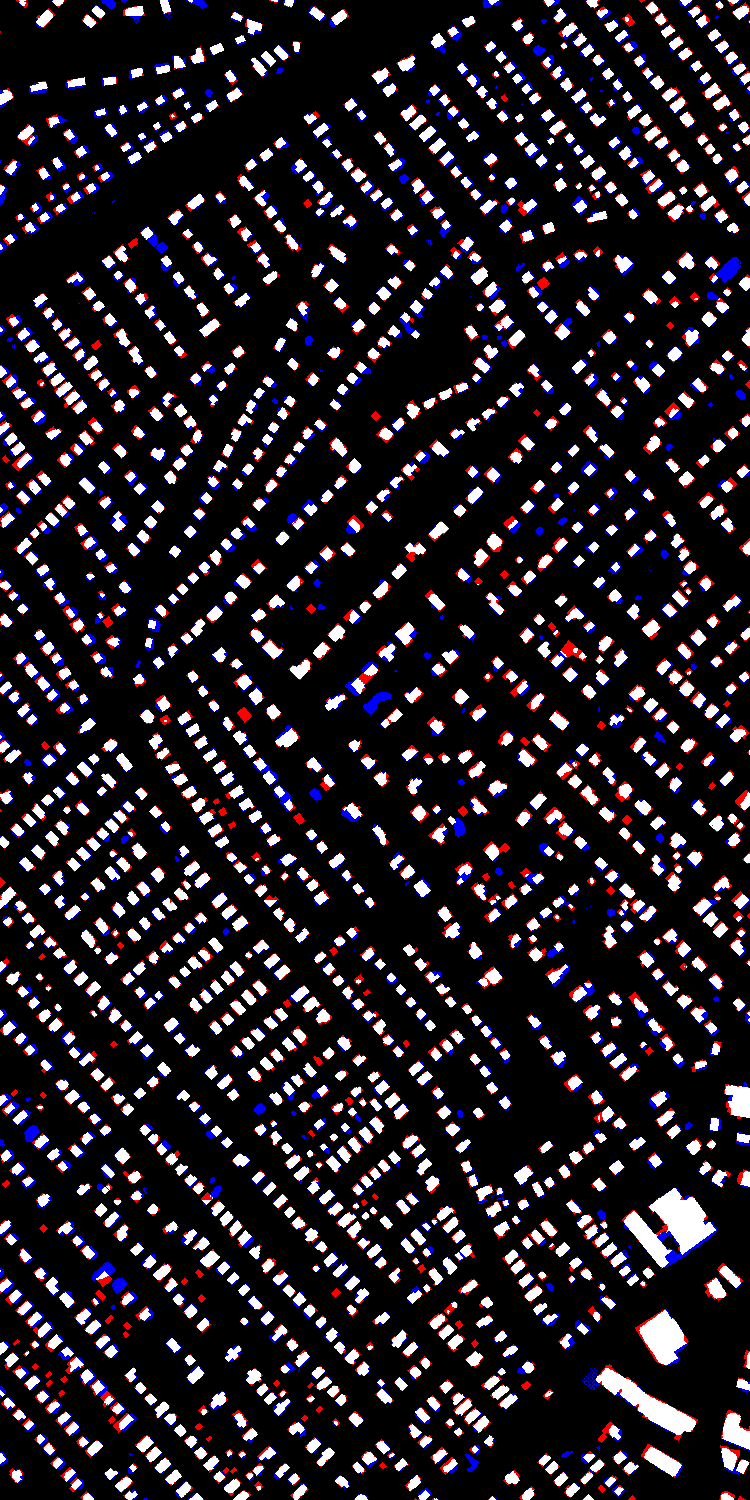}}

    \caption{Building segmentation results using our proposed method on Massachusetts building dataset. The colors white, black, blue and red indicate true positives, true negatives, false positives and false negatives, respectively.}
    \label{fig:intro_picture}
\end{figure}

Many of the earlier approaches relied on hand-engineered features for building extraction. They exploited the features such as the structure, color and hyper-spectral data of remote sensing images to improve performance. These feature-based methods are coupled with machine learning algorithms for detection and classification \cite{forlani2006complete, frontoni2008comparative, ok2008robust}. However, due to the limitations of low-level features, these algorithms have low performance. In contrast to traditional methods, deep learning methods benefit from learning features by optimizing an objective function. The success of deep learning algorithms, such as Convolutional Neural Networks (CNNs), has resulted in an improved performance on various computer vision tasks. These advances in deep learning have benefited several remote sensing applications, such as aerial image object detection, image retrieval.

Building detection in remote sensing is usually posed as a segmentation problem. Recent works have obtained state-of-the-art results in semantic segmentation for remote sensing imagery. Most works have explored encoder-decoder network structures to limit the parameter increase in the bottleneck layers for semantic segmentation \cite{long2015fully, jegou2017one, ronneberger2015u}. Fully Convolutional Networks (FCNs) introduced the first encoder-decoder structure for semantic segmentation. FCNs replaced the fully connected layers with a fully convolutional layer, which reduces the number of parameters of the CNN model \cite{long2015fully}. Other work has built further on this encoder-decoder structure, while improving segmentation performance.

Apart from utilizing a sophisticated architecture, most CNN-based segmentation algorithms rely on cross-entropy or similar loss functions as the objective function. Due to the limitations imposed by cross-entropy, we explore an alternative loss function inspired by adversarial learning, that preserves structure and refines the results without the need of additional post-processing steps like Conditional Random Fields (CRFs). In particular, we explore adversarial learning in conjunction with direct per-pixel optimization utilized by cross-entropy loss \cite{goodfellow2014generative}.

\section{Related Work}
Satellite imagery has been systematically captured over the last decade and a large amount of research has been conducted by both the remote sensing and the computer vision communities. Several approaches have been proposed for segmentation of buildings and other terrestrial objects from aerial imagery. 

\paragraph{Image segmentation}: In recent years, deep learning algorithms have provided state-of-the-art results for segmentation. Earlier work that relied on deep learning used fully connected layers to produce a vector that was later reshaped to a tensor \cite{mnih2010learning, saito2015building, sebastian2018bootstrapped}.
However, this has been replaced by fully convolutional layers which have removed the output size restriction. Most segmentation networks use an encoder-decoder architecture such as a Fully Convolutional Network (FCN), U-Net, etc. However, the unique nature of remote sensing imagery has fuelled the design of several custom architectures which have been also proposed for aerial image segmentation~\cite{ronneberger2015u}. Much of the work on building segmentation has been focused on improving the network architecture. The success of deep learning in computer vision has also resulted in concentrating on network architectures specifically designed for aerial image segmentation. Approaches such as attention \cite{huang2019automatic}, larger receptive fields \cite{liu2018semantic}, and other post-processing techniques are often added to existing networks to improve aerial image segmentation \cite{liu2018semantic}. Besides these aspects, a few works have also considered larger context as input for the networks \cite{DBLP:journals/corr/Marcu16, sebastian2018bootstrapped}. Context provides an understanding of the object inside and provides higher quality segmentation. However, most of these works consider per-pixel loss to improve the performance, rather than capturing the properties of aerial imagery.

\paragraph{Adversarial learning} 
Adversarial learning has been primarily explored for generative models, where it is used to synthesize perceptually realistic images \cite{goodfellow2014generative}. Adversarial learning is also used to create robust models against adversarial attacks. An adversarial learning approach utilizes a discriminator network besides the generator network, to distinguish real and fake samples. Instead of post-processing techniques such as CRFs, adversarial learning provides conditioning and structure to the segmentation outputs. Few previous publications have considered adversarial learning for semantic segmentation \cite{luc2016semantic}. The general approach in the previous results deploys a pair-wise input to the discriminator, where both the generated and input images are fed to the discriminator. Finally, the direct pixel-level and adversarial losses are combined in a weighted scheme. Adding the adversarial loss fills in missing regions by learning to capture the overall structure of a building, similar to inpainting to task of from context information~\cite{sebastian2018conditional}.

\begin{figure*}[t]
\begin{center}
   \includegraphics[width=1.0\linewidth]{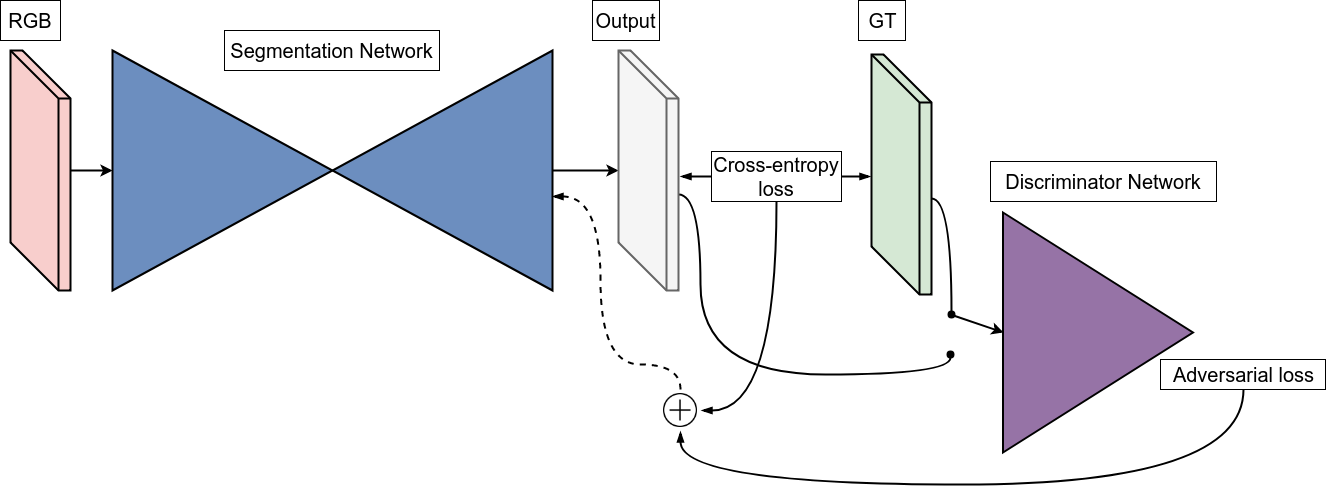}
\end{center}
   \caption{Overview of the proposed method. Dotted lines indicate combined cross entropy and adversarial loss of the generator samples, fed back to update the segmentation network parameters. The adversarial loss of both real and generated samples are used to update the discriminator parameters (not shown in this figure).}
\label{fig:pipeline}
\end{figure*} 

\section{Method}
\subsection{Dataset}
For the experiments, we use the Massachusetts building dataset \cite{MnihThesis}. It consists of 151~high-resolution RGB aerial images of regions in Boston. Each image has a resolution of 1,500 $\times$ 1,500 pixels with a spatial resolution of one square meter per pixel. The regions depict primarily urban and sub-urban areas with a coverage of 340~m{$^2$}. The dataset is split into~137, 4~and 10~images for training, validation and testing. During training, each image is divided into 300 $\times$ 300 pixel patches without any overlap. Data augmentation is performed by flipping the images left-right and top-down and applying rotations of 90\degree, 180\degree and 270\degree. 

\subsection{Adversarial learning}
We combine both adversarial and cross-entropy losses to jointly optimize the generator. The final loss $\mathcal{L}$ is defined as 
\begin{equation}
\mathcal{L} = 
\underset{\textit{G}}{\textnormal{min }} \underset{\textit{D}}{\textnormal{max }} \mathcal{L_{\textnormal{adv}}} (\textit{G}, \textit{D}) + \mathcal{L_{\textnormal{c.e}}} (\textit{G}),
\end{equation}
\noindent
where \textit{G} is the generator network and \textit{D} is the discriminator network. Parameter $\mathcal{L_{\textnormal{adv}}}$ is the adversarial loss and  $\mathcal{L_{\textnormal{c.e}}}$ is the cross-entropy loss. Losses $\mathcal{L_{\textnormal{adv}}}$ and $\mathcal{L_{\textnormal{c.e}}}$ are specified by

\begin{equation}
	\begin{aligned}
		\mathcal{L_{\textnormal{adv}}} (\textit{G}, \textit{D})  = 
		\underset{y\sim \mathbb{P}_{l}} 
		{\mathbb{E}}[\text{log}(\textit{D}(y))] + 
		\underset{x \sim \mathbb{P}_{r}} 
		{\mathbb{E}}[\text{log}(1 - \textit{D}(G({x})))] ,
	\end{aligned}
\end{equation}

\begin{equation}
    \mathcal{L}_{c.e}(G) = - y\cdot \log(\hat{y}) + (1 - y)\cdot \log(1 -\hat{y}),
\end{equation}
\noindent
 where $\hat{y}$ denotes the output from the segmentation network $\textit{G(x)}$, parameter $y$ is the label (sampled from a real distribution $\mathbb{P}_{l}$) and $x$ (sampled from a real distribution $\mathbb{P}_{r}$) is the input RGB image. During training, the generator weights are updated based on the combination of equally weighted adversarial and cross-entropy losses.

\subsection{Network architectures}
The segmentation network acts as a generator for adversarial training. Like most adversarial training approaches, our architecture is composed of a generator and a discriminator. An overview of the method is shown in Figure~\ref{fig:pipeline}.

\subsubsection{Generator}
To test the effectiveness of adversarial learning in conjunction with cross-entropy loss, we deploy the combined loss on several existing state-of-the-art networks. We test the new loss on DeepLab~v3+ \cite{chen2018encoder}, DenseNet \cite{huang2016densely, jegou2017one}, and PSPNet \cite{zhao2017pyramid}. DeepLab~v3+ is an extension to the series of DeepLab architectures. DeepLab~v3+ consists of Atrous Spatial Pyramid Pooling, combined with the low-level features from earlier layers of a pre-trained ResNet model~\cite{he2015deep}. DenseNet is comprised of dense blocks where every layer is connected to every other layer by concatenation. PSPNet utilizes a pooling module where the final output of the network is pooled at different levels and are combined via concatenation. 

\subsubsection{Discriminator}
In all experiments, we employ the same discriminator architecture. Unlike previous work, we do not have a symmetric discriminator as the generator. Our discriminator consists of 4~convolution layers (3 $\times$ 3 kernel size with~32, 64, 128, 256~filters) and 2~fully connected layers (512, 1~outputs) that classify the image as real or fake. Batch normalization~\cite{ioffe2015batch} is not applied and exponential linear units are used for training the discriminator as done in~\cite{ghafoorian2018gan}. 

\subsection{Implementation details}
All the networks that are trained with cross-entropy loss using the Adam optimizer ($\beta_1$=0.9 and $\beta_2$=0.99) with a batch size of~3 for 90~epochs. Both DeepLab~v3+ and PSPNet networks are trained using a pretrained model with a learning rate of 10${^{-5}}$. DenseNet is trained without pretraining with a learning rate of 10${^{-4}}$. The networks trained with adversarial and cross-entropy losses use the same settings as above for the generator (the segmentation network). The discriminator is also trained with the Adam optimizer ($\beta_1$=0.5 and $\beta_2$=0.9) with a learning rate of 10${^{-6}}$ for DeepLab~v3+ and PSPNet, and 10${^{-5}}$ for DenseNet. The discriminator to generator training ratio is set to unity. 

\subsection{Evaluation metric}
The commonly used metrics for the evaluation of detection results are the precision and recall measures. Precision and recall are also known as correctness and completeness in remote sensing literature. For evaluation, we use the Accuracy, the $F_1$ measure and the mean IoU (mIoU) metrics, to obtain valid comparisons with previous work. Accuracy is computed by
\begin{equation}
    \text{Accuracy} = \frac{T.P + T.N}{T.P + T.N + F.P + F.N},
\end{equation}
\noindent
while the $F_1$ score and Precision and Recall are defined by 

\begin{equation}
    F_1 = 2 \cdot \frac{\text{precision} \cdot \text{recall}}{\text{precision} + \text{recall}},
\end{equation}

 \begin{equation}
    \text{Precision}  = \frac{T.P}{T.P + F.P} , \hspace{5mm}   
    \text{Recall} = \frac{T.P}{T.P + F.N} .
 \end{equation}

\begin{figure*}[t]
\captionsetup[subfloat]{labelformat=empty}
    \centering
    \subfloat{\includegraphics[width=0.1667\linewidth]{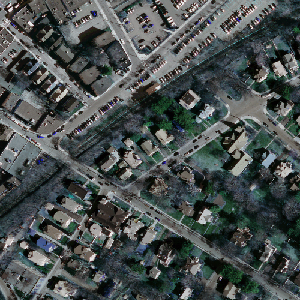}}
    \subfloat{\includegraphics[width=0.1667\linewidth]{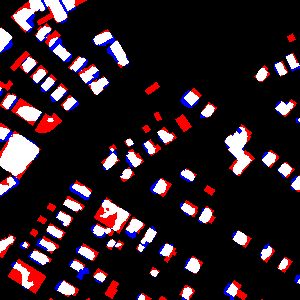}}
    \subfloat{\includegraphics[width=0.1667\linewidth]{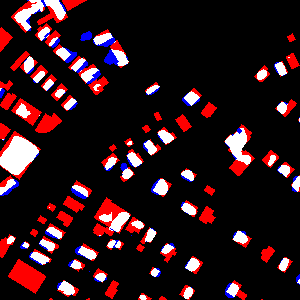}}
    \subfloat{\includegraphics[width=0.1667\linewidth]{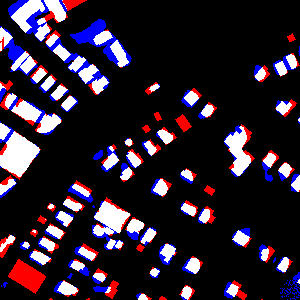}}
    \subfloat{\includegraphics[width=0.1667\linewidth]{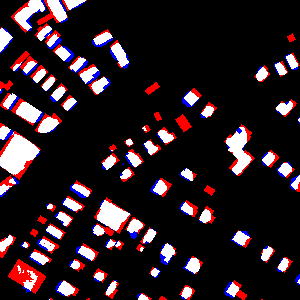}}
    \subfloat{\includegraphics[width=0.1667\linewidth]{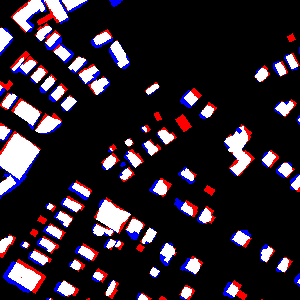}}
    \vspace{-0.82\baselineskip}
    \subfloat{\includegraphics[width=0.1667\linewidth]{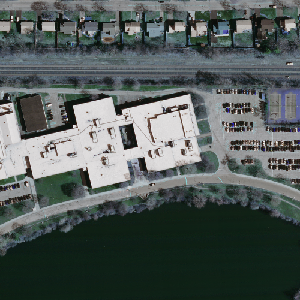}}
    \subfloat{\includegraphics[width=0.1667\linewidth]{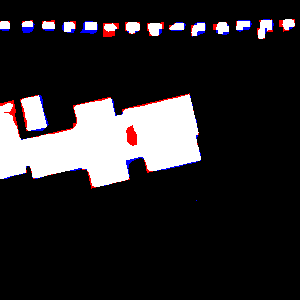}}
    \subfloat{\includegraphics[width=0.1667\linewidth]{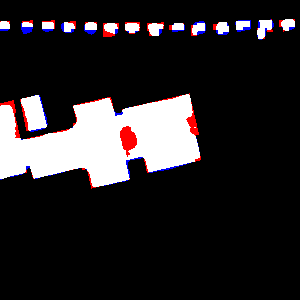}}
    \subfloat{\includegraphics[width=0.1667\linewidth]{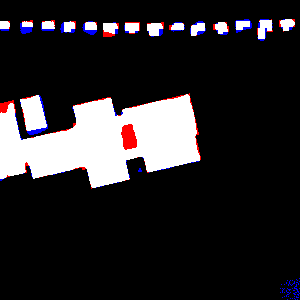}}
    \subfloat{\includegraphics[width=0.1667\linewidth]{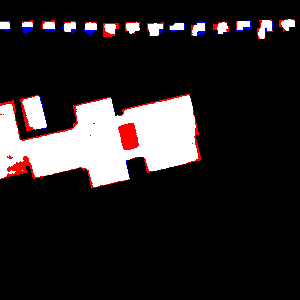}}
    \subfloat{\includegraphics[width=0.1667\linewidth]{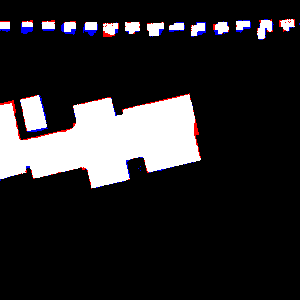}}
    \vspace{-0.82\baselineskip}
    \subfloat{\includegraphics[width=0.1667\linewidth]{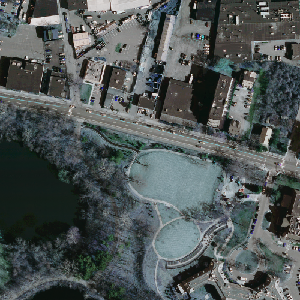}}
    \subfloat{\includegraphics[width=0.1667\linewidth]{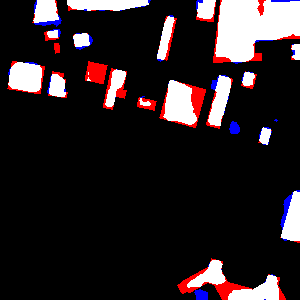}}
    \subfloat{\includegraphics[width=0.1667\linewidth]{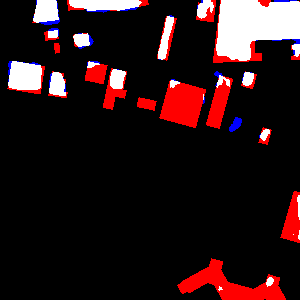}}
    \subfloat{\includegraphics[width=0.1667\linewidth]{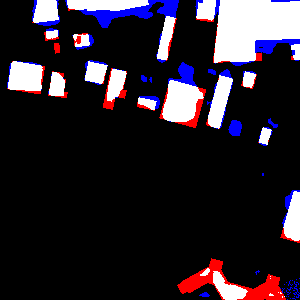}}
    \subfloat{\includegraphics[width=0.1667\linewidth]{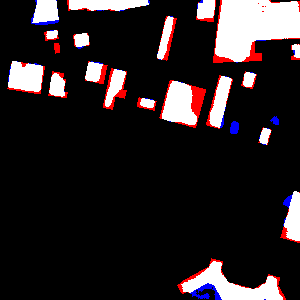}}
    \subfloat{\includegraphics[width=0.1667\linewidth]{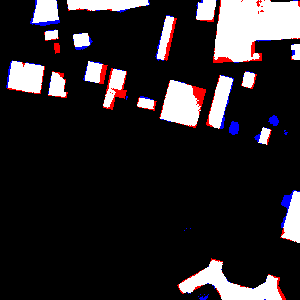}}
    \vspace{-0.82\baselineskip}
    \subfloat{\includegraphics[width=0.1667\linewidth]{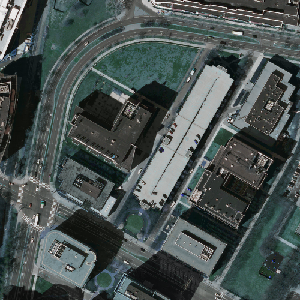}}
    \subfloat{\includegraphics[width=0.1667\linewidth]{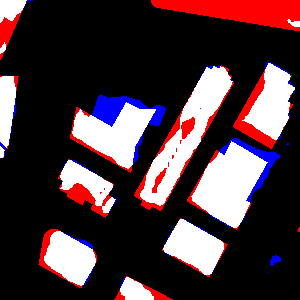}}
    \subfloat{\includegraphics[width=0.1667\linewidth]{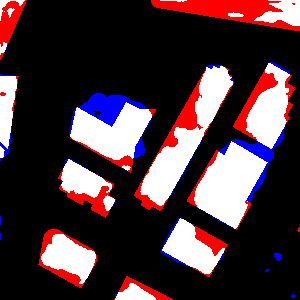}}
    \subfloat{\includegraphics[width=0.1667\linewidth]{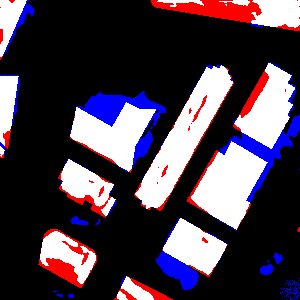}}
    \subfloat{\includegraphics[width=0.1667\linewidth]{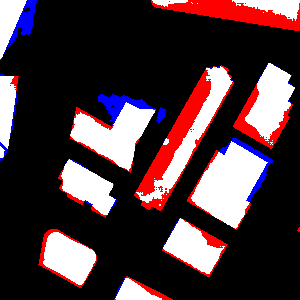}}
    \subfloat{\includegraphics[width=0.1667\linewidth]{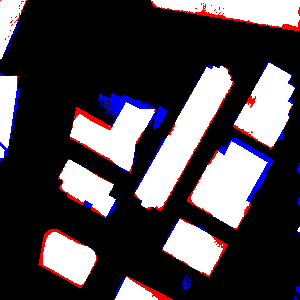}}
    \vspace{-0.82\baselineskip}
    \subfloat[Aerial Image]{\includegraphics[width=0.1667\linewidth]{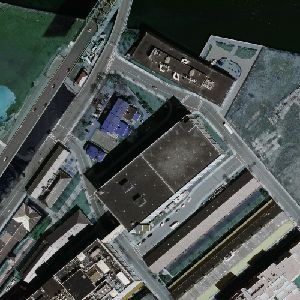}}
    \subfloat[DeepLab v3+]{\includegraphics[width=0.1667\linewidth]{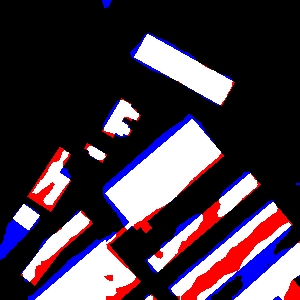}}
    \subfloat[PSPNet]{\includegraphics[width=0.1667\linewidth]{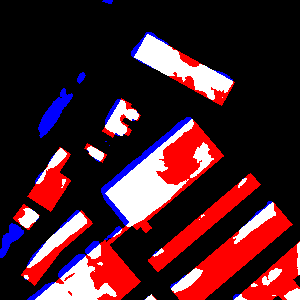}}
    \subfloat[PSPNet + adv.]{\includegraphics[width=0.1667\linewidth]{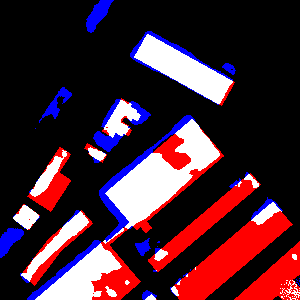}}
    \subfloat[FC-DenseNet]{\includegraphics[width=0.1667\linewidth]{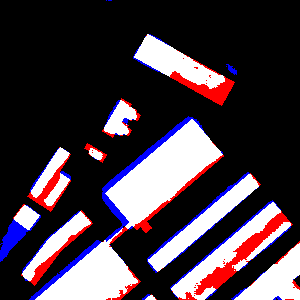}}
    \subfloat[FC-DenseNet + adv.]{\includegraphics[width=0.1667\linewidth]{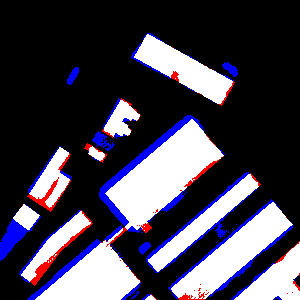}}
    \caption{Building segmentation results for various segmentation networks. The +adv indicates the addition of adversarial loss with cross-entropy loss. The colors white, black, blue and red pixels indicate true positives, true negatives, false positives and false negatives, respectively.}
    \label{fig:Inpaintingexamples}

\end{figure*}
\noindent
Similary, IoU is computed as 
\begin{equation}
    \text{IoU} = \frac{T.P}{T.P + F.P + F.N},
\end{equation}
where $T.P$, $T.N$, $F.P$, and $F.N$ are the true positives, true negatives, false positives, false negatives, respectively.
In building and road detection, the relaxed version of the $F_1$ and IoU metrics are used. The relaxed version of Precision is the fraction of predicted building pixels that are within a radius of $\rho$ pixels of the ground-truth building pixel, whereas the relaxed Recall represents the fraction of ground-truth building pixels that are within $\rho$ pixels of a predicted building pixel. The value of $\rho$ is set to $\rho$=3 in all the experiments, which is identical to previous work.

\begin{table}
\begin{center}
   \caption{Performance of different methods on the Massachusetts Building dataset. Best results are presented in bold, second best are between [ ] brackets. Results in ( ) parenthesis are improvements with adversarial + cross-entropy loss.}
\begin{tabular}{|l|l|l|l|}
\hline
\textbf{Network}         & \textbf{Accuracy}      & \textbf{Relaxed F1}    & \textbf{Relaxed IoU}    \\ \hline
Mnih \& Hinton           & -                      & 92.11                  & -                       \\ \hline
Saito et al.             & -                      & [94.88]                & -                       \\ \hline
ELU-FCN-CRF              & -                      & 93.93                  & 89.08                   \\ \hline
Dual Path Network        & -                      & 94.23                  & -                       \\ \hline
DeepLab v3+              & 92.13                  & 92.65                  & 86.31                   \\ \hline
PSPNet                 & 90.9                   & 89.52                  & 81.2                      \\ \hline
PSPNet + adv.          & 91.02 (+0.12)          & 91.17 (+1.65)          & 83.78 (+2.58)             \\ \hline
FC-DenseNet            & [93.18]                & 94.33                  & [89.27]                   \\ \hline
FC-DenseNet + adv      & \textbf{93.45 (+0.27)} & \textbf{95.59 (+1.26)} & \textbf{91.55 (+2.28)}    \\ \hline
\end{tabular}
\end{center}
   \label{table:results}
\end{table}

\section{Results}
To compare the results against previous methods, we measure the performance across different metrics. The results are summarized in Table~1. From this table, it is evident that Fully Convolutional DenseNet (FC-DenseNet) with adversarial loss offers the best performance. We observe that with sufficient data augmentation, we are able to produce competitive results with other state-of-the-art methods. The addition of adversarial loss to the segmentation task consistently offers better performance across all the metrics. We observe this positive trend with both DenseNet and PSPNet. Note that the first 3~methods in Table~1 deploy CRFs as a post-processing step. Compared to dual-path networks, our method is significantly less expensive to train and run inference, since dual-path networks use parameter intensive AlexNet and VGGNet to learn global and local features. During inference, the discriminator network is not used and hence has the same computation cost of running a standard segmentation network. From qualitative results, it can be seen that the addition of the adversarial loss fills in missing regions more coherently than the standard cross-entropy loss. This effect is visualized in Figure~\ref{fig:Inpaintingexamples}. The proposed system partly acts as an inpainting network where a prior segmentation is generated simultaneously using the cross-entropy loss and is reconstructed by the adversarial loss.

\section{Conclusion}
We have proposed a loss function to train CNNs for semantic segmentation of aerial imagery. The proposed loss function, which is a combination of the adversarial and cross-entropy losses, consistently improves performance without any additional cost during inference. We have concluded that the addition of the adversarial loss improves the overall structure and produces a more coherent output taking  the  context into consideration. Furthermore, our method has been evaluated across commonly used metrics and a comparison with state-of-the-art methods is provided. Finally, the proposed method outperforms the state-of-the-art results on the Massachusetts building dataset with a relaxed $F_1$ of 95.59\% without any additional post-processing techniques.

{\small
\bibliographystyle{IEEEtran}
\bibliography{ref}
}

\end{document}